\documentclass{ifacconf}
\usepackage{lipsum}
\usepackage{dsfont}
\usepackage{cite}
\usepackage{amsmath,amssymb,amsfonts}
\usepackage{algorithm}

\usepackage{algorithmic}
\usepackage{graphicx}
\usepackage{textcomp}
\usepackage{xcolor}

\def\BibTeX{{\rm B\kern-.05em{\sc i\kern-.025em b}\kern-.08em
    T\kern-.1667em\lower.7ex\hbox{E}\kern-.125emX}}
\usepackage{graphicx}      
\usepackage{natbib}        
\newtheorem{secthm}{Theorem}[section]
\newtheorem{seccor}[secthm]{Corollary}
\newtheorem{seclem}[secthm]{Lemma}

\newtheorem{secprob}[secthm]{Problem}
\newtheorem{secdefn}[secthm]{Definition}
\newtheorem{secrem}[secthm]{Remark}

\newcommand{\rank}{{\operatorname{rank}}}

\newcommand{\bE} { {\mathbb E}}
\newcommand{\bP} { {\mathbb P}}
\newcommand{\bR} { {\mathbb R}}
\newcommand{\bS} { {\mathbb S}}
\newcommand{\bZ} { {\mathbb Z}}

\newcommand{\cF} { {\mathcal F}}

\newcommand{\cN} { {\mathcal N}}
\newcommand{\esssup}{\operatorname*{ess\,sup}}
\renewcommand{\theenumi}{\roman{enumi}}

\def\red{\hfill $\lhd$}

\begin{document}
\begin{frontmatter}

\title{A Randomized Scheduling Framework for Privacy-Preserving Multi-robot Rendezvous given Prior Information\thanksref{footnoteinfo}} 
\thanks[footnoteinfo]{The work of Liu and Cao was supported in part by the Talent Program (Vici-19901) and the ELSA Lab for Technical Industry (ELSA4TI) of the Netherlands Organization for Scientific Research. The work of Liu is partly supported by China Scholarship Council. The work of Yu Kawano was supported in part by JSPS KAKENHI Grant Number 25K22804.}

\author[First]{Le Liu} 
\author[Second]{Yu Kawano} 
\author[First]{Ming Cao}

\address[First]{Faculty of Science and Engineering, University of Groningen, 9747 AG Groningen, The Netherlands, (e-mail: le.liu@rug.nl; m.cao@rug.nl).}
\address[Second]{Graduate School of Advance Science and Engineering, Hiroshima University, Higashi-Hiroshima 739-8527, Japan, (e-mail: ykawano@hiroshima-u.ac.jp)}

\begin{abstract}                Privacy has become a critical concern in modern multi-robot systems, driven by both ethical considerations and operational constraints. As a result, growing attention has been directed toward privacy-preserving coordination in dynamical multi-robot systems. This work introduces a randomized scheduling mechanism for privacy-preserving robot rendezvous. The proposed approach achieves improved privacy even at lower communication rates, where privacy is quantified via pointwise maximal leakage. We show that lower transmission rates provide stronger privacy guarantees and prove that rendezvous is still achieved under the randomized scheduling mechanism. Numerical simulations are provided to demonstrate the effectiveness of the method.
\end{abstract}

\begin{keyword}
Privacy, Robot rendezvous, Consensus, Random scheduling, Bayesian inference
\end{keyword}

\end{frontmatter}
\section{Introduction}
The growing interconnection of cyber–physical systems has enabled unprecedented collaboration and efficiency across modern industrial applications~\citep{han2018privacy}. From smart grids and intelligent transportation to cooperative robots, these distributed systems rely on the exchange of, sometimes sensitive, information among multiple agents~\citep{hassan2019differential}. Such information sharing, however, introduces significant privacy risks that may lead to security vulnerabilities or violations of data protection regulations~\citep{wang2024survey}. In the multi-robot rendezvous problem, where agents must meet at a common location, each robot’s initial state or position can reveal sensitive or proprietary information, such as deployment strategies or operational patterns that adversaries or competitors could exploit~\citep{yu2024panav, chatzimichali2020toward}. Protecting these sensitive states is therefore a critical privacy requirement in practical rendezvous scenarios.

In recent years, numerous privacy-preserving mechanisms have been proposed for multi-robot coordination and rendezvous problems (e.g.,~\citep{nozari2017differentially, mo2016privacy, altafini2019dynamical}). Approaches using differential privacy and cryptographic techniques have been explored to limit information leakage~\citep{Yu2020}. However,  efficient communication is critical in the robot rendezvous problem, as practical multi-robot systems often operate under restricted transmission rates. This observation raises an intriguing question: can these inherent communication constraints be leveraged to enhance privacy in rendezvous tasks? Motivated by this insight, this paper investigates the interplay between communication efficiency and privacy preservation in the robot rendezvous problem, aiming to develop mechanisms that achieve both objectives simultaneously.

\subsection{Literature Review}
Privacy preservation in distributed and multi-robot  systems has gained significant attention in recent years due to the increasing need to protect sensitive information in networked control and cyber-physical systems. Early works primarily adopted differential privacy (DP) as a formal framework for quantifying privacy leakage~\citep{liu2023privacy, liu2025privacy, liu2025initial, liu2025quantization}.
For example, \citet{liu2023privacy} introduced a differentially private tracking control algorithm using stochastic quantizers. Building on this idea, \citet{liu2025quantization} further developed dynamic quantizers to balance control performance and privacy.
Differential privacy for initial states on a Riemannian manifold is addressed in~\citep{liu2025initial}. In addition, \citet{nozari2017differentially} established rigorous privacy guarantees for distributed average consensus while preserving convergence.

Beyond noise-based methods, several works \citep{ruan2019secure,altafini2019dynamical} have explored encryption strategies. For example, \citet{ruan2019secure} examined encryption-based consensus and ensure accurate consensus. An accurate consensus mechanism that preserves privacy was proposed in \citep{altafini2019dynamical} based on a dynamical masking approach.

{Although extensive research has studied privacy in dynamical systems, the practical case where prior information about robots' initial locations is publicly available has received limited attention. 
To address this issue, we adopt PML as the privacy metric. 
Unlike differential privacy, PML explicitly incorporates the adversary's prior knowledge. At the same time, PML provides a pointwise worst-case measure of posterior-to-prior amplification. 
This makes PML more suitable for our adversarial model, where the adversary has prior knowledge of each robot's initial state distribution and observes the released output history.}

\subsection{Contribution}
To address the challenge of achieving both communication efficiency and initial-state privacy in robot rendezvous when robots’ prior information is publicly known, we employ the pointwise maximal leakage (PML)~\citep{liu2025privacyprotectionexposuresystems} framework and introduce a randomized scheduling mechanism for privacy-preserving coordination. The proposed method randomizes the communication schedule to limit the exposure of each robot’s information, thereby enhancing initial-state privacy while simultaneously reducing the overall communication rate. Privacy is quantified using the PML metric, which captures the worst-case information gained by an adversary observing the communication pattern. We theoretically show that reducing the communication rate leads to improved privacy, and further prove that rendezvous can still be achieved under the proposed mechanism.

\subsection{Organization}
The remainder of this paper is organized as follows. Section~\ref{sec:pre} formulates the privacy-preserving rendezvous problem, introduces the randomized scheduling strategy, and defines the PML privacy metric. Section~\ref{sec:results} presents the privacy analysis and proves that rendezvous can be achieved under the proposed random scheduling mechanism. Section~\ref{sec:sim} illustrates the effectiveness of the proposed method through a numerical simulation. Finally, Section~\ref{sec:con} concludes the paper with some remarks.
\subsection{Notation}
The sets of real numbers and non-negative integers are denoted by $\bR$ and  $\mathbb{Z}_+$, respectively.  Let $\cN(\mu, \Sigma)$ represent a Gaussian distribution with the mean $\mu \in \bR^n$ and covariance $\Sigma \in \bS_{++}^{n}$, where $\bS_{++}^{n}$ denotes the set of $n \times n$ symmetric and positive definite matrices. A probability space is denoted by $(\Omega, \cF, \bP )$, where $\Omega$, $\cF$, and $\bP $ denote the sample space, $\sigma$-algebra, and probability measure, respectively. The expectation of a random variable is denoted by $\bE[\cdot]$.

Consider a communication network represented by a graph $\mathcal{G} = (\mathcal{V}, E)$, 
where $\mathcal{V} = \{1, 2, \ldots, N\}$ denotes the set of robots, and 
$E\subseteq \mathcal{V} \times \mathcal{V}$ represents the set of communication links.  
Let $A = [a_{ij}] \in \mathbb{R}^{N \times N}$ denote the weighted adjacency matrix, 
where $a_{ij} > 0$ if $(i, j) \in E$ and $a_{ij} = 0$ otherwise. We do not allow self-loops so that $a_{ii} = 0, \forall i \in \mathcal{V}$.
The degree matrix is defined as $D = \mathrm{diag}(d_1, d_2, \ldots, d_N)$, 
where each diagonal entry is given by $d_i = \sum_{j=1}^N a_{ij}$. 
Throughout this paper, we assume that the graph $\mathcal{G}$ is undirected and connected, and $\sum_{j=1}^N a_{ij} < 1$ for all $i \in \mathcal{V}$.

\section{Problem Formulation}
\label{sec:pre}
In many practical rendezvous scenarios, a robot’s initial position encodes sensitive operational meaning. For example, in multi-company inspection of shared industrial facilities, robots start the rendezvous protocol from locations tied to ongoing inspection tasks or maintenance priorities. Information exchanged during rendezvous can inadvertently reveal these positions, exposing fault-prone components or strategic operational decisions. An adversary observing the protocol could thus infer weaknesses or resource allocations. Consequently, safeguarding the privacy of initial robot positions is essential in real-world multi-robot rendezvous settings. Motivated by this, our goal in this paper is to protect the privacy of initial robot locations while ensuring successful rendezvous.
\subsection{Privacy Preserving Robot Rendezvous Problem}
Consider a multi-robot system operating in a two-dimensional space, with an undirected and connected communication graph $\mathcal{G}$, described by
\begin{align}
\label{eq:sys_ori}
    x_i(k+1)=x_i(k)+u_i(k), \forall i \in \mathcal{V},
\end{align}
where $x_i(k) \in \mathbb{R}^2$ and  $u_i(k) \in \mathbb{R}^2$ denote the location and velocity of the robot $i$ to be controlled. The initial state $x_i(0)$ is drawn from a public Gaussian prior $\mathcal{N}(\mu_i,r_i I_2)$.
Such prior knowledge arises naturally in many settings, e.g., from historical data or physical constraints. 
When prior information is public, classical differential privacy is \emph{not} an ideal criterion because differential privacy does not explicitly account for prior information and can therefore be either ineffective or conservative~\citep{saeidian2023pointwise_tit}. Accordingly, we adopt an alternative privacy measure that directly incorporates the prior; this criterion is introduced and formalized later.

Our objective is to study a privacy preserving rendezvous problem. Namely, our goal is to design $u_i(k)$ such that
\begin{align*}
    \lim_{k \to \infty} ( x_{i}(k) - x_{j}(k) ) = 0, 
    \quad \forall i,j \in \mathcal{V},
\end{align*}
while simultaneously preventing others from inferring $x_i(0)$ using communication data. Before proceeding to the controller design, we first introduce a random scheduling strategy, which helps privacy protection.

\subsection{Random Scheduling}
\label{sec:random_scheduling}
The random scheduling is a mechanism in which each node independently decides whether to transmit data at each time instant according to a tunable probability \(p_i \in (0,1]\). This simple modification serves two goals: it reduces aggregate communication load while improving privacy. 

For notational clarity, the random scheduling scheme is encoded using an indicator function.
Let $\gamma_i(k)\in\{0,1\}$ indicate whether robot $i$ transmits data at time $k$, i.e.,
\begin{align}
\label{eq:ber}
    \gamma_{i}(k)= \begin{cases}1, & \text { robot } i \text { transmits data  at time $k$, } \\ 0, & \text { otherwise, }\end{cases}
\end{align}
where 
\begin{align*} 
\begin{cases}\mathbb{P}\!\left(\gamma_i(k)=1\right)=p_i \in (0,1], \\ \mathbb{P}\!\left(\gamma_i(k)=0\right)=1-p_i,
\end{cases}
\end{align*}
and $\gamma_{i}(k)$ are independent across $i$ and $k$.

To limit privacy leakage when $\gamma_i(k)=1$, robot $i$ adds noise to its message. Specifically, it transmits the perturbed value $x_i(k) + v_i(k)$, where $v_i(k) \sim \mathcal{N}(0,\sigma_i^2(k) I_2)$ is Gaussian noise, independent across robots and time, also independent of $\gamma_j(k)$ for all $j \in \mathcal{V}$ and $k \in \bZ_{+}$.

When $\gamma_i(k)=0$, robot $i$ does not transmit the value $x_i(k)$ at time $k$, and the other robots update using the most recent available value from $i$. 
For notational brevity, we define the following variable that captures the recent available output from robot $i$:
\begin{align}
\label{eq:recent}
    \tilde{x}_i(k) =
    \begin{cases}
       x_i(k) + v_i(k), &\text{ if }  \gamma_{i}(k) = 1, \\
       \tilde{x}_{i}(k-1), &  \text{ if } \gamma_{i}(k) = 0,
    \end{cases}
\end{align}
where $\tilde{x}_{i}(-1) = \mu_i$. Then, the control protocol $u_i(k)$ can be designed as
\begin{align*}
    u_i(k) = \sum_{j =1}^{N} a_{ij}(\tilde{x}_j (k) - x_i(k)).
\end{align*}
Therefore, the system dynamics for robot $i$ can be written as
\begin{align}
\label{eq:sys}
    x_{i}(k+1) = x_{i}(k) + \sum_{j =1}^{N} a_{ij}(\tilde{x}_j (k) - x_i(k)).
\end{align}
Note that exact average rendezvous is achieved when $v_i(k)=0$ for all $i \in \mathcal{V},\ k \in \mathbb{Z}_{+}$, while $p_i=1$ for all $i \in \mathcal{V}$ under the assumption that $\mathcal{G}$ is undirected and connected, and $\sum_{j=1}^N a_{ij} < 1$ for all $i \in \mathcal{V}$.
Under the random scheduling mechanism introduced in this section, exact average rendezvous is generally impossible, and even the rendezvous itself may not be ensured.

\subsection{Pointwise Maximal Leakage}
In this subsection, we formalize the privacy notion used in this paper. We adopt pointwise maximal leakage (PML) as our privacy criterion, as it explicitly accounts for adversaries with prior knowledge of each robot’s initial state and captures their worst-case inference capability~\citep{saeidian2023pointwise, liu2025privacyprotectionexposuresystems}. To rigorously formulate the problem, we first define the output sequence that collects all outputs of robot $i$ up to time $t$ as
\begin{align*}
    \tilde{x}_i(0{:}t)
    := 
    \begin{bmatrix}
        \tilde{x}_i(0) & \tilde{x}_i(1) & \cdots & \tilde{x}_i(t)
    \end{bmatrix}.
\end{align*}
The indicator variables up to time $t$ are denoted by
\begin{align*}
    \gamma_{i}(0{:}t) = \begin{bmatrix}
        \gamma_i(0) & \gamma_i(1) &\cdots & \gamma_i(t)
    \end{bmatrix}.
\end{align*}
In this paper, we consider an adversary who is capable of eavesdropping on all outputs generated by all robots. Consequently, the information accessible to the adversary up to time $t$ is
\begin{align*}
    \tilde{x}(0{:}t) = \begin{bmatrix}
        \tilde{x}_1(0{:}t)^{\top} & \tilde{x}_2(0{:}t)^{\top} &\cdots & \tilde{x}_N(0{:}t)^{\top}
    \end{bmatrix}^{\top}
\end{align*}
and
\begin{align*}
    \gamma(0{:}t) = \begin{bmatrix}
        \gamma_1(0{:}t)^{\top} & \gamma_2(0{:}t)^{\top} &\cdots & \gamma_N(0{:}t)^{\top}
    \end{bmatrix}^{\top}.
\end{align*}

Given the adversary’s information 
\begin{align*}
\mathcal{I}(t):= \begin{bmatrix}
    \tilde{x}(0{:}t)^{\top} & \gamma(0{:}t)^{\top}
\end{bmatrix}^{\top}, 
\end{align*}
the formal definition of PML is as follows,

\begin{secdefn} \label{def:pml_metric}
    The {\emph{pointwise maximal leakage} (PML)} from $x_i(0)$ to $\mathcal{I}(t)$ is defined by
    \begin{align}
    \label{eq:PML}
        \ell (x_i(0) \rightarrow \mathcal{I}(t))  := \log \underset{x_i(0) \in \bR^2}{\esssup} \frac{f(x_i(0) \mid \mathcal{I}(t))}{f(x_i(0))},
    \end{align}
    where $f(x_i(0) \mid \mathcal{I}(t))$ stands for the posterior distribution under $\mathcal{I}(t)$, $f(x_i(0))$ is the probability density function of publicly known prior distribution~$\mathcal{N}(\mu_i, r_i I_2)$, and $\esssup$ denotes the supremum up to prior-measure-zero sets.
 \red
\end{secdefn}

Informally, PML measures how much the posterior deviates from the prior—i.e., the maximal information gain from observing the output history. 
A large value of $\ell\!\left(x_i(0)\to\mathcal{I}(t)\right)$ indicates that the posterior has shifted substantially relative to the prior, which is undesirable from a privacy standpoint. 
Accordingly, we intend to keep $\ell\!\left(x_i(0)\to\mathcal{I}(t)\right)$ small for strong privacy guarantees. To capture the failure probability, we introduce the following $(\varepsilon,\delta)$-PML privacy, which is the privacy criterion we use in this paper.

\begin{secdefn}
\label{def:pml_privacy}
Given $\varepsilon_i \geq 0$ and $\delta_i \in [0,1]$, the system~\eqref{eq:sys} is said to be \emph{$(\varepsilon_i, \delta_i)$-PML private} if
\begin{align}
\label{def:PML}
    \bP [ \ell (x_i(0) \rightarrow \mathcal{I}(t)) \leq \varepsilon_i] \geq 1-\delta_i, \forall t \in \bZ_{+}
\end{align}
holds, where the probability is taken with respect to the randomness of $(\gamma_i(k),\, v_i(k))$ and the PML $\ell (x_i(0) \rightarrow \mathcal{I}(t)) $ is defined by \eqref{eq:PML}.
\red
\end{secdefn}

Similar to differential privacy~\citep{dwork2006differential,dwork2006calibrating}, smaller values of $(\varepsilon_i,\delta_i)$ in PML privacy correspond to stronger privacy guarantees. {Although our framework can be defined in a finite-time setting (i.e., protecting privacy from finite data), this paper studies PML privacy over an infinite time horizon since $t$ is arbitrary in~\eqref{eq:PML}.} This choice models the most powerful eavesdropper and is consistent with prior studies (e.g.,~\cite{mo2016privacy,nozari2017differentially}), which evaluate privacy over an infinite time horizon.

\begin{secrem}
    {
In contrast to the settings in~\citep{mo2016privacy,nozari2017differentially}, this paper considers a randomized scheduling framework, where each robot transmits intermittently with a tunable probability. 
This mechanism reduces communication load and also limits privacy leakage. 
Unlike differential privacy considered in~\cite{nozari2017differentially}, PML incorporates the adversary's prior knowledge and can therefore provide a more tailored privacy characterization for our setting. It also differs from the estimation-based metric in~\cite{mo2016privacy}, which evaluates privacy through estimation accuracy rather than the information revealed by the observations.
\red
}
\end{secrem}

\begin{secprob}
    In this paper, we are interested in answering the following two problems.
    \begin{itemize}
        \item How should the random scheduling probabilities \(p_i\) and the noise variance \(\sigma_i^2(k)\) be chosen to guarantee \((\varepsilon,\delta)\)-PML at each time \(k\) ?
        \item Can the robots reach rendezvous while protecting privacy at the same time?
        \red
    \end{itemize} 
\end{secprob}
\section{Main Results}
\label{sec:results}
In this section, we present the main results. First, we derive conditions on the scheduling probabilities $p_i$ and noise variances $\sigma_i^2(k)$ under which the system~\eqref{eq:sys} is \((\varepsilon,\delta)\)-PML private. Next, under this calibrated choice of $(p_i,\sigma_i^2(k))$, we prove that rendezvous is achieved. All the proofs are provided in Appendices.

\subsection{Privacy Analysis}
As evident from the definition of PML, both $\varepsilon$ and $\delta$ depend on $(p_i, \sigma_i^2(k))$. However, obviously, arbitrary choices of $(p_i, \sigma_i^2(k))$ cannot guarantee rendezvous, even though privacy can be preserved. To address this, we first propose our privacy-preserving algorithm in Algorithm~\ref{alg:rand-schedule}. Subsequently, the following result establishes a closed-form relationship between $(\varepsilon_i, \delta_i)$ and $(p_i, \sigma_i^2(k))$ under Algorithm~\ref{alg:rand-schedule}.

\begin{algorithm}[t]
\caption{Privacy-Preserving rendezvous via Random Scheduling}
\label{alg:rand-schedule}
\begin{algorithmic}[1]
  \REQUIRE Symmetric weights $a_{ij} = a_{ji} \ge 0$ with $d_i = \sum_{j=1}^N a_{ij} < 1$; 
           transmission probabilities $p_i \in (0,1]$; 
           noise scales $\sigma_i > 0$; 
           decay factors $q_i \in (0,1)$; 
           initial states $x_i(0)$ for all $i = 1,\dots,N$.
  \STATE Initialize $\tilde{x}_i(-1) \gets \mu_i$ for all $i = 1,\dots,N$.
  \FOR{$k = 0,1,2,\dots$}
    \FOR{$i = 1$ to $N$}
      \STATE Draw $\gamma_i(k) \sim \mathrm{Bernoulli}(p_i)$, independently across $i$ and $k$ and independently of the past.
      \IF{$\gamma_i(k) = 1$}
        \STATE Set $\sigma_i^2(k) \gets q_i^{2k} \sigma_i^2$.
        \STATE Draw $v_i(k) \sim \mathcal{N}\big(0,\sigma_i^2(k) I_2\big)$, independently across $i$ and $k$ and independently of $\{\gamma_j(\ell)\}$.
        \STATE $\tilde{x}_i(k) \gets x_i(k) + v_i(k)$.
      \ELSE
        \STATE $\tilde{x}_i(k) \gets \tilde{x}_i(k-1)$.
      \ENDIF
    \ENDFOR
    \FOR{$i = 1$ to $N$}
      \STATE $u_i(k) \gets \displaystyle \sum_{j=1}^{N} a_{ij}\big(\tilde{x}_j(k) - x_i(k)\big)$.
      \STATE $x_i(k+1) \gets x_i(k) + u_i(k)$.
    \ENDFOR
  \ENDFOR
\end{algorithmic}
\end{algorithm}

\begin{secthm}
\label{thm:main}
     Given $(\varepsilon_i, \delta_i)$ in Algorithm~\ref{alg:rand-schedule},
suppose the design parameters $(p_i, \sigma_i^2, q_i)$ are selected to satisfy
\begin{align*}
    \alpha_i := 1 - d_i < q_i < 1,\qquad
\varepsilon_i \ge \log\!\left(1 + \frac{r_i}{\sigma_i^2(1-\rho_i)}\right),
\end{align*}
where $\rho_i := \alpha_i^{2}/q_i^{2}$. Then, the system~\eqref{eq:sys} is $(\varepsilon_i, \delta_i)$-PML private if 
      \begin{align}
      \label{eq:thm_main}
          \delta_i \geq 1-\sum_{k=0}^{\infty}(1-p_i)^k p_i b_k(\varepsilon_i, \sigma_i^2,q_i),
      \end{align}
      where  $b_k(\varepsilon_i, \sigma_i^2,q_i) := \mathbf{F}_{\chi^2_2}\big(2\varepsilon_i - 2\log (1 + \frac{r_i \rho_i^k}{\sigma_i^2(1-\rho_i)})\big)$, and $\mathbf{F}_{\chi^2_2}$ denotes cumulative distribution function of a $\chi^2$ distribution with 2 degrees of freedom.
\end{secthm}
\begin{pf}
The proof can be found in the full version of this paper in~\cite{liu2025rendezvous}.
\qed
\end{pf}

Our decaying noise strategy is similar to that in~\citep{nozari2017differentially}. However, due to existence of $\gamma_i$, it is not yet clear whether rendezvous can be achieved.
Before discussing how to design the random scheduling mechanism to make the system~\eqref{eq:sys} $(\varepsilon,\delta)$-PML private, we provide several remarks on Theorem~\ref{thm:main}.

\begin{secrem}
    First, by setting $p_i = 1$, we recover the noise--only case, which coincides with the decaying noise strategy of \citep{nozari2017differentially}.
Second, when $\sigma_i$, $q_i$, and $\varepsilon_i$ are fixed, the privacy parameter $\delta_i$ is an increasing function of $p_i$. In fact, if we define
\begin{align*}
    \Phi_i(p_i) = 1 - \sum_{k=0}^{\infty}(1-p_i)^k p_i b_k(\varepsilon_i, \sigma_i^2,q_i),
\end{align*}
one can calculate 
\begin{align*}
    &\Phi_i(1) - \Phi_i(p_i) \\
    =& \sum_{k=0}^{\infty}(1-p_i)^k p_i 
    \big(b_k(\varepsilon_i, \sigma_i^2,q_i) 
    - b_0(\varepsilon_i, \sigma_i^2,q_i) \big) \\
    \geq & p_i(1-p_i)
    \big(b_1(\varepsilon_i, \sigma_i^2,q_i) 
    - b_0(\varepsilon_i, \sigma_i^2,q_i) \big) \geq 0,
\end{align*}
where we have used the convention that $0^0 = 1$ for calculating $\Phi_i(1)$.
This observation confirms that a lower transmission rate enhances privacy.
\end{secrem} 

    \begin{secrem}
{The parameter $q_i$ determines how fast the injected noise decays. 
A larger $q_i$ improves privacy by keeping the noise effective for longer, but it may slow down rendezvous convergence. 
Thus, in practice, $q_i$ should be chosen as the smallest value that satisfies the desired $(\varepsilon_i,\delta_i)$-PML bound in Corollary~\ref{cor:design}, so that privacy is guaranteed while the impact on convergence is limited.}
\end{secrem}

Since Theorem~\ref{thm:main} involves an infinite sum, it is not convenient for practical use. Instead, it motivates a constructive guideline for designing the random scheduling mechanism. This result is summarized in the next corollary.
\begin{seccor} 
\label{cor:design}
Given $(\varepsilon_i, \delta_i)$ in Algorithm~\ref{alg:rand-schedule},
suppose the design parameters $(p_i, \sigma_i^2, q_i)$ are selected  such that $\alpha_i < q_i < 1$ and
\begin{align}
\label{eq:design}
    \varepsilon_i \geq \log\!\left(1+\frac{r_i}{\sigma_i^2(1-\rho_i)}\right) + \frac{1}{2}\mathbf{F}_{\chi^2_2}^{-1}(1-\tilde{\delta}_i),
\end{align}
where $\tilde{\delta}_i = \delta_i + (1-p_i)\big(b_1(\varepsilon_i, \sigma_i^2,q_i) - b_0(\varepsilon_i, \sigma_i^2,q_i) \big)$. Then, the system~\eqref{eq:sys} is $(\varepsilon_i,\delta_i)$-PML private.
\end{seccor}
\begin{pf}
    The proof can be found in the full version of this paper in~\cite{liu2025rendezvous}.
    \qed
\end{pf}

One may first select $(\sigma_i^2, q_i)$ such that the system~\eqref{eq:sys} is $(\varepsilon_i, \tilde{\delta}_i)$-PML private by enforcing the condition obtained from~\eqref{eq:thm_main} with $p_i = 1$. Then the system~\eqref{eq:sys} is $(\varepsilon_i, \delta_i)$-PML private with
\begin{align*}
\delta_i \;=\; \tilde{\delta}_i - (1-p_i)\big(b_1(\varepsilon_i, \sigma_i^2, q_i) - b_0(\varepsilon_i, \sigma_i^2, q_i)\big).
\end{align*}
This again demonstrates that the random transmission rate $p_i$ can further amplify privacy.

Although we have established sufficient conditions for the system~\eqref{eq:sys} to achieve $(\epsilon_i,\delta_i)$-PML privacy, the corresponding rendezvous under the random scheduling scheme remains unclear. This issue is addressed in the next subsection.

\subsection{Rendezvous Analysis}
In this subsection, we provide the result confirming the rendezvous property under Algorithm~\ref{alg:rand-schedule}.
\begin{secthm}
\label{thm:con}
    The system~\eqref{eq:sys} achieves rendezvous under Algorithm~\ref{alg:rand-schedule}, i.e.,
    \begin{align*}
        \lim_{k \to \infty} (x_i(k) - x_j(k))= 0 \; \mathrm{ a.s.} \quad \forall\, i,j\in \mathcal{V}.
    \end{align*}
\end{secthm}
\begin{pf}
    The proof can be found in the full version of this paper in~\cite{liu2025rendezvous}.
    \qed
\end{pf}

The proof of Theorem~\ref{thm:con} relies on constructing an appropriate Lyapunov candidate that quantifies the disagreement among robots and then showing that its conditional expectation satisfies a contraction-type inequality. Due to the random scheduling mechanism introduced in Section~\ref{sec:random_scheduling}, the recent location updates of the neighbor robots occur according to the random scheduling. Consequently, the analysis can only guarantee convergence in the almost-sure sense. In contrast, under a deterministic communication scheme, rendezvous can typically be established deterministically. 

\begin{secrem}
{The proposed randomized scheduling and noise-injection mechanism inevitably introduces performance loss. 
First, the injected noise may cause the final rendezvous point to deviate from the exact average of the initial states. 
Second, random scheduling reduces the frequency of communication and may therefore slow down the convergence process. 
In this paper, we focus on showing that rendezvous and PML privacy can be simultaneously achieved under the proposed framework. 
A quantitative characterization of the trade-off among privacy level, convergence rate, communication burden, and rendezvous accuracy is an important direction for future work.
\red}
\end{secrem}

\section{Numerical Experiments}
\label{sec:sim} 
{In this section, we illustrate the proposed design using a simple numerical example. 
We first specify the desired privacy requirement, then present the associated randomized scheduling and noise-injection scheme, and finally demonstrate the resulting rendezvous performance.}

We require each robot to satisfy $(\varepsilon_i,\delta_i)$-PML privacy with
\[
    \varepsilon_i = 5,\qquad \delta_i = 0.015,\qquad i\in\mathcal{V}.
\]
The network consists of $N=5$ robots with the following undirected weight matrix:
\[
A \!=\! \begin{bmatrix}
0    & 0.3 & 0.3 & 0 & 0\\
0.3 & 0    & 0.2 & 0 & 0\\
0.3 & 0.2 & 0    & 0.3 & 0\\
0 & 0 & 0.3 & 0    & 0.5\\
0 & 0 & 0 & 0.5 & 0
\end{bmatrix}
\]
and the initial states are independently sampled from
\[
x_i(0)\sim \mathcal{N}(\mu_0,\sigma_0^2 I_2),
\qquad \forall i \in \mathcal{V},
\]
where $\mu_0=0,\sigma_0^2=3$.
According to Corollary~\ref{cor:design}, we choose the randomized scheduling probabilities as
\[
p = [\,0.6,\ 0.6,\ 0.6,\ 0.6,\ 0.6\,]^\top,
\]
and the noise parameters as
\begin{align*}
    \sigma = [\,1,\ 1,\ 1,\ 1,\ 1\,]^\top, 
    q = [\,0.9,\ 0.9,\ 0.9,\ 0.9,\ 0.9\,]^\top.
\end{align*}

\begin{figure}
    \centering
    \includegraphics[width=1\linewidth]{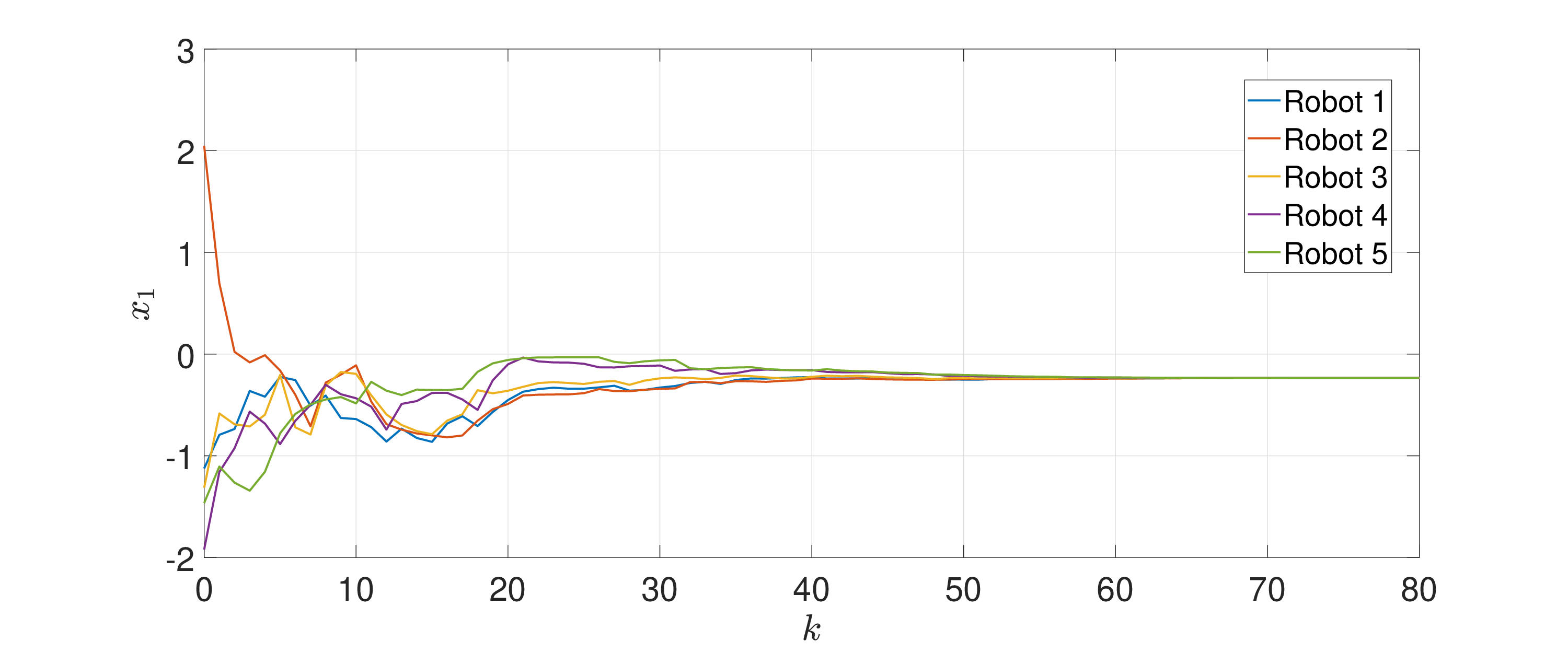}
    \caption{State trajectories of the first coordinate.}
    \label{fig:consensus1}
\end{figure}

\begin{figure}
    \centering
    \includegraphics[width=1\linewidth]{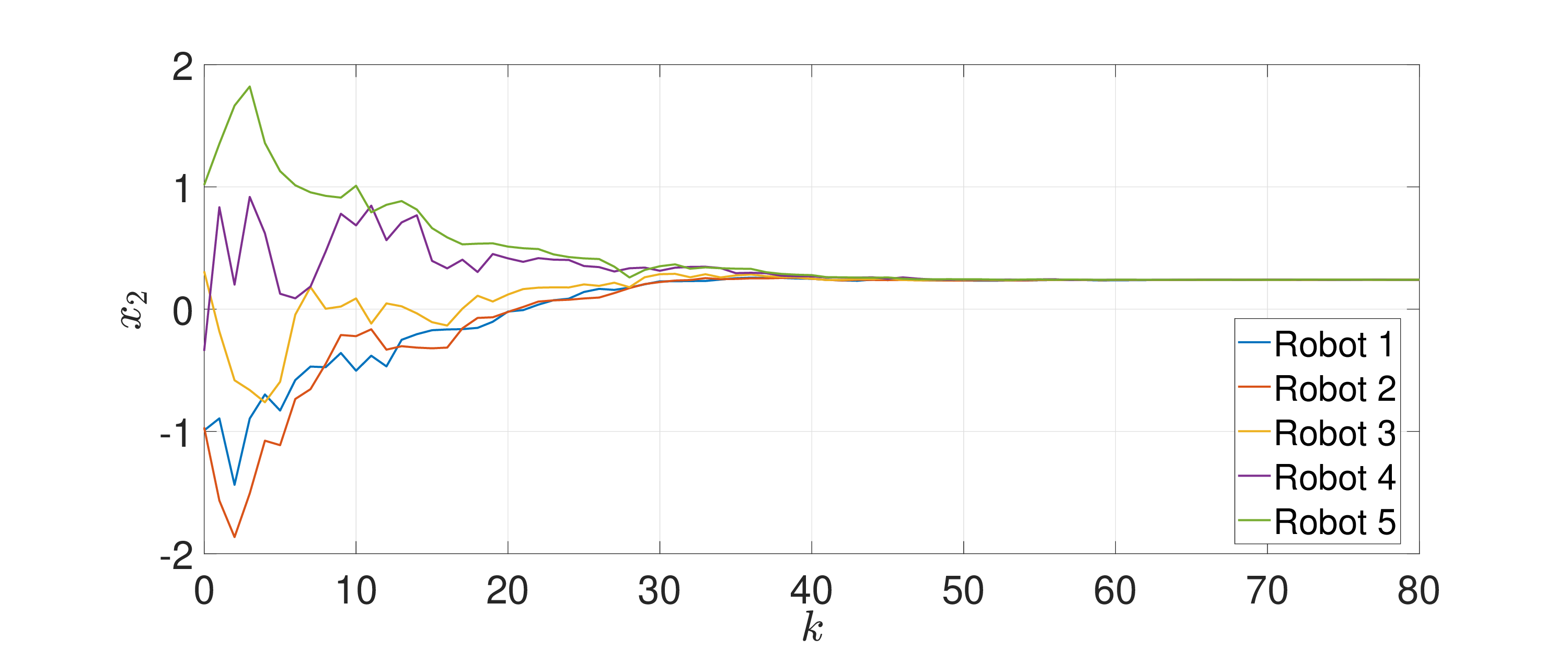}
    \caption{State trajectories of the second coordinate.}
    \label{fig:consensus2}
\end{figure}

The simulation results are shown in Figs.~\ref{fig:consensus1} and~\ref{fig:consensus2}.

{To highlight the role of randomized scheduling, we note that the proposed scheme differs from existing privacy-preserving consensus approaches such as~\cite{nozari2017differentially}, where agents communicate at every iteration and the communication probability is not a design variable. 
Therefore, a direct comparison would not isolate the contribution of randomized scheduling, which in our framework is used to jointly enhance privacy and reduce communication load. }

As shown in Figs.~\ref{fig:consensus1} and~\ref{fig:consensus2}, all robot states reach rendezvous under the designed randomized scheduling and noise-injection scheme. 
Together with the privacy parameters computed above, this demonstrates that the proposed method can simultaneously achieve the prescribed $(5,0.015)$-PML privacy requirement, reduce communication load, and preserve rendezvous performance.
\section{Conclusion}
\label{sec:con}
In this paper, a random scheduling mechanism has been proposed to achieve privacy-preserving rendezvous in multi-robot systems. By adopting pointwise maximal leakage as the privacy metric, it has been shown that reducing the communication rate enhances privacy while still ensuring rendezvous convergence. We have established a sufficient condition for achieving rendezvous under the proposed scheme, and a numerical simulation has verified the effectiveness of the method. Since the final rendezvous location under the random scheduling strategy does not equal the true average, this results in a performance loss. Future work includes investigating privacy–performance trade-offs.
\appendix

\section{Proof of Theorem~\ref{thm:main}}
\label{app:1}
To present the proof, we first provide the following lemma, which gives a closed form of PML for a linear Gaussian mechanism.
\begin{seclem}
\label{lem:basic}
       Suppose $X$ follows the Gaussian distribution $\mathcal{N}(\mu_X, \Sigma_X)$. The PML from $X \in \bR^n$ to $Y \in \bR^m$ for a linear Gaussian mechanism $Y = CX + V$ satisfies
    \begin{align}
    \label{eq:pml_linear}
       \ell(X \to Y) = \frac{1}{2}\log \det (I_{n} + \Sigma_{X}C^{\top} R^{-1}C)+\frac{1}{2} \xi,
    \end{align}
    where $V \sim \mathcal{N}(0, R)$, $\xi$ follows a $l$-freedom $\chi^2$ distribution, and $l = \rank(C)$.  
\end{seclem}
\begin{pf}
    The proof can be directly completed by the result established in \citep[Theorem 3.2]{liu2025privacyprotectionexposuresystems}.
    \qed
\end{pf}

{\bf Proof of Theorem~\ref{thm:main}.}
    We compute the PML $\ell (x_i(0) \rightarrow \mathcal{I}(t))$ with $\mathcal{I}(t): = [\begin{matrix}\tilde{x}(0{:}t)^{\top} & \gamma(0{:}t)^{\top}\end{matrix}]^\top$ by first transforming the outputs into their innovation form, since innovation sequences provide a clean characterization of the input--output relationship involving $x_i(0)$. 
For each robot~$i$, define the innovation sequence as
\begin{align*}
z_i(0) :=& \gamma_{i}(0)\tilde{x}_i(0), \\
    z_i(k) 
    :=& \gamma_{i}(k)\left(\tilde{x}_i(k) - \sum_{s=0}^{k-1} \alpha_i^{\,k-1-s} 
       \sum_{j=1}^N a_{ij}\, \tilde{x}_j(s)\right),
     \forall k \geq 1,
\end{align*}
where $\alpha_i$ is defined in Theorem~\ref{thm:main}. 
Let 
\begin{align*}
    z_i(0{:}t) :=& 
\begin{bmatrix}
    z_i(0) & \cdots & z_i(t)
\end{bmatrix}^{\!\top},
\\
z(0{:}t) :=& 
\begin{bmatrix}
    z_1(0{:}t)^{\top} & \cdots & z_N(0{:}t)^{\top}
\end{bmatrix}^{\top}.
\end{align*}
{
Next, we show that the joint information vectors 
$\tilde{\mathcal{I}}(t) := [z(0{:}t)^{\top},\, \gamma(0{:}t)^{\top}]^{\top}$ and $\mathcal{I}(t)$ 
can be recovered from each other through causal mappings. 
From the definition of $z(0{:}t)$, it is immediate that $\tilde{\mathcal{I}}(t)$ can be constructed from $\mathcal{I}(t)$. 
We now establish the reverse direction. 
First, observe that $\tilde{x}_i(0) = \tilde{x}_i(-1)$ if $\gamma_i(0) = 0$, and $\tilde{x}_i(0) = z_i(0)$ if $\gamma_i(0) = 1$.
For \(k \ge 1\), the reconstruction proceeds by induction. 
Assume that \(\tilde{x}(0),\dots,\tilde{x}(k-1)\) have already been recovered. 
Then the quantity $\sum_{s=0}^{k-1}\alpha_i^{\,k-1-s}\sum_{j=1}^N a_{ij}\tilde{x}_j(s)$
is known. 
Using the definition of \(z_i(k)\), we obtain
\begin{align*}
\tilde{x}_i(k)=
\begin{cases}
z_i(k)+\displaystyle\sum_{s=0}^{k-1}\alpha_i^{\,k-1-s}\sum_{j=1}^N a_{ij}\tilde{x}_j(s), & \gamma_i(k)=1,\\[1mm]
\tilde{x}_i(k-1), & \gamma_i(k)=0.
\end{cases}
\end{align*}
Thus, \(\tilde{x}_i(k)\) can always be recovered using only \(z_i(k)\), \(\gamma_i(k)\), and past values of \(\tilde{x}\).}
Hence, the posterior distributions conditioned on $\mathcal{I}(t)$ and $\tilde{\mathcal{I}}(t)$ are identical, and the corresponding PML are equivalent in the sense that
\begin{align*}
\ell\big(x_i(0)\!\rightarrow\!\mathcal{I}(t)\big)
    = \ell\big(x_i(0)\!\rightarrow\!\tilde{\mathcal{I}}(t)\big).
\end{align*}
{From~\eqref{eq:sys}, one can verify by induction that
\begin{align*}
x_i(k)=\alpha_i^{\,k}x_i(0)
+\sum_{s=0}^{k-1}\alpha_i^{\,k-1-s}
\sum_{j=1}^N a_{ij}\tilde{x}_j(s), \forall k \geq 1
\end{align*}
Under the random scheduling mechanism~\eqref{eq:recent}, if $\gamma_i(k)=1$, substituting $\tilde{x}_i(k)=x_i(k)+v_i(k)$ yields
\begin{align*}
    z_i(k)=\alpha_i^{\,k}x_i(0)+v_i(k).
\end{align*}
If $\gamma_i(k)=0$, then $z_i(k)=0$. 
Hence, for all $k\in\mathbb{Z}_+$,
\begin{align*}
z_i(k)=\gamma_i(k)\big(\alpha_i^{\,k}x_i(0)+v_i(k)\big).
\end{align*}}
Then, it can be summarized that
\begin{equation}
\label{eq:innovation}
    z_i(k)
    = \gamma_i(k)\big(\alpha_i^{\,k} x_i(0) + v_i(k)\big),
    \forall k \in \mathbb{Z}_{+}.
\end{equation}
Therefore, only the innovation sequence $z_i(0{:}t)$ is statistically correlated with $x_i(0)$. 
It follows that
\begin{align*}  \ell\big(x_i(0)\!\rightarrow\!\tilde{\mathcal{I}}(t)\big)
    = \ell\big(x_i(0)\!\rightarrow\!z_i(0{:}t)\big).
\end{align*}
Furthermore, the innovation vector $z_i(0{:}t)$ can be rewritten as
\begin{align*}
    z_i(0{:}t) = C_i(\gamma_i(0{:}t),t)x_i(0) + \tilde{v}(t),
\end{align*}
where
\begin{align*}
C_i(\gamma_i(0{:}t),t)
:= \begin{bmatrix}
\gamma_i(0) I_2 \\
\gamma_i(1)\alpha_i I_2 \\
\vdots \\
\gamma_i(t)\alpha_i^t I_2
\end{bmatrix},
\qquad
\tilde{v}(t)
:= \begin{bmatrix}
\gamma_i(0)v_i(0)^{\top} &
\gamma_i(1)v_i(1)^{\top} &
\cdots &
\gamma_i(t)v_i(t)^{\top}
\end{bmatrix}^{\top}.
\end{align*}
Leveraging Lemma~\ref{lem:basic} gives 
\begin{align*}
    \ell(X \to Z_i(0{:}t)) = \log\left(1 + \sum_{k=0}^{t} r_i \gamma_i(k)\frac{\alpha_i^{2k}}{\sigma_{i}^2(k)}\right)+\frac{1}{2} \xi,
\end{align*}
where $\xi$ follows an $l$-freedom $\chi^2$ distribution and $l = \rank( C_i(\gamma_i(0{:}t),t))$.

Recall that $\sigma_i^2(k) = \sigma^2_i q_i^{2k}$, $\rho_i = \frac{\alpha_i^2}{q_i^2}$ and use the fact that $\rank( C_i(\gamma_i(0{:}t),t)) \leq 2$, it further follows that
\begin{align*}
    \ell(X \to Z_i(0{:}t)) \le \log \left(1 + \sum_{k=0}^{\infty}\gamma_i(k)\frac{r_i}{\sigma_i^2}\rho_i^{k}\right)+\frac{1}{2} \tilde{\xi},
\end{align*}
where $\tilde{\xi}$ follows a $2$-freedom $\chi^2$ distribution.

Then, to prove the system~\eqref{eq:sys} is $(\varepsilon_i,\delta_i)$-PML private, it suffices to ensure that 
\begin{align}
\label{eq:proof_thm1}
    \mathbb{P}\left[\log \left(1 + \sum_{t=0}^{\infty}\gamma_i(t) \frac{r_i}{\sigma_i^2} \rho_i^{t}\right)+\frac{1}{2} \tilde{\xi} \leq \varepsilon_i\right] \geq 1-\delta_i.
\end{align}
Let $K=\min \left\{k \geq 0: \gamma_i(k)=1\right\}$ be the index of the first transmission.
Then, the events $\{K=k\}$ are disjoint with
\begin{align*}
    \mathbb{P}(K=k)=(1-p_i)^k p_i, k \in \bZ_{+}.
\end{align*}
On the event $\{K = k\}$, we have $\gamma_i(t) = 0$ for $t < k$ and $\gamma_i(t) \le 1$ for $t \ge k$. Hence, it follows that
\begin{align}
    \label{eq:thm1_4}
\sum_{t=0}^{\infty}\gamma_i(t) \frac{r_i}{\sigma_i^2} \rho_i^{t} \leq \frac{r_i \rho_i^k}{\sigma_i^2(1-\rho_i)}
\end{align}
under the event $\{K = k\}$.
Therefore, leveraging the fact that $\mathbf{F}_{\chi^2_2}$ is an increasing function and~\eqref{eq:thm1_4}, we have
\begin{align*}
&\mathbb{P}\!\left[\log\!\Big(1 + \sum_{t=0}^{\infty}\gamma_i(t)\, \frac{r_i}{\sigma_i^2}\rho_i^{t}\Big)
      + \tfrac{1}{2}\tilde{\xi} \le \varepsilon_i \right] \\
&= \mathbb{E}\!\left[
      \mathbb{P}\!\left(
          \log\!\Big(1 + \sum_{t=0}^{\infty}\gamma_i(t)\, \frac{r_i}{\sigma_i^2}\rho_i^{t}\Big)
          + \tfrac{1}{2}\tilde{\xi} \le \varepsilon_i \,\Big|\, K
      \right)
   \right] \\
&= \sum_{k=0}^{\infty} \mathbb{P}(K = k)
\mathbb{E}\!\left[
\mathbf{F}_{\chi^2_2}\!\left(
2\varepsilon_i - 2\log\!\Big(1 + \sum_{t=0}^{\infty}\gamma_i(t)\,
\frac{r_i}{\sigma_i^2}\rho_i^{t}\Big)
\right)
\,\middle|\, K = k
\right] \\
&\ge \sum_{k=0}^{\infty} (1-p_i)^k p_i\,
      \mathbf{F}_{\chi^2_2}\left(
          2\varepsilon_i - 2\log\!\Big(1 + \tfrac{r_i \rho_i^k}{\sigma_i^2(1-\rho_i)}\Big)
      \right) \\
&\ge 1 - \delta_i.
\end{align*}
where the last inequality is from~\eqref{eq:thm_main}.
This completes the proof.
    \qed

\section{Proof of Corollary~\ref{cor:design}}
\label{app:2}
By~\eqref{eq:design} and the monotonicity of $\mathbf{F}_{\chi^2_2}(\cdot)$, it follows that
\begin{align*}
    \tilde{\delta}_i \geq 1 - b_0(\varepsilon_i, \sigma_i^2, q_i).
\end{align*}
Then, recalling that
\begin{align*}
    \tilde{\delta}_i
= \delta_i
+ (1-p_i)\big(
b_1(\varepsilon_i, \sigma_i^2, q_i)
- b_0(\varepsilon_i, \sigma_i^2, q_i)
\big),
\end{align*}
we obtain
\begin{align*}
    \delta_i \geq& 1 - b_0(\varepsilon_i, \sigma_i^2, q_i) - (1-p_i)\big(
b_1(\varepsilon_i, \sigma_i^2, q_i)
- b_0(\varepsilon_i, \sigma_i^2, q_i)
\big) \\
= & 1-\big(p_i b_0\left(\varepsilon_i, \sigma_i^2, q_i\right)+\left(1-p_i\right) b_1\left(\varepsilon_i, \sigma_i^2, q_i\right)\big) \\
\geq & 1-\sum_{k=0}^{\infty}(1-p_i)^k p_i b_k(\varepsilon_i, \sigma_i^2,q_i),
\end{align*}
where we have used that $b_k(\varepsilon_i, \sigma_i^2, q_i)$ is increasing in $k$ and $\sum_{k=1}^{\infty}(1-p_i)^k p_i b_k \ge (1-p_i) b_1$.
Therefore, the system~\eqref{eq:sys} is $(\varepsilon_i,\delta_i)$-PML private by Theorem~\ref{thm:main}.
\qed

\section{Proof of Theorem~\ref{thm:con}}
\label{app:3}
The proof is divided into three steps:
(i) We formulate an augmented system to facilitate the stability proof of system~\eqref{eq:sys} and define the corresponding Lyapunov candidate;
(ii) we show that the Lyapunov candidate decreases on an event determined by the random scheduling indicators $\gamma_i(k)$, and remains non-increasing at all other events;
(iii)  we conclude that rendezvous is achieved almost surely.

(i) For each agent $i$, define the augmented state 
$   \eta_i(k) := \begin{bmatrix} x_i(k) & \tilde x_i(k) \end{bmatrix}^{\top} \in \mathbb{R}^{4} $,
$
\eta(k) := [\eta_1(k)^\top,\dots,\eta_N(k)^\top]^\top \in \mathbb{R}^{4N}.
$ and 
$\Gamma(k):=\mathrm{diag}(\gamma_1(k),\dots,\gamma_N(k))$. 
The closed-loop dynamics~\eqref{eq:recent} and~\eqref{eq:sys} can be written compactly as
\begin{align*}
    \eta(k{+}1)
= M(\Gamma(k+1))\,\eta(k)
+ N(\Gamma(k+1))\,v(k+1),
\end{align*}
where
\[
M(\Gamma)=
\begin{bmatrix}
I_n-D & A\\
\Gamma(I_n-D) & \Gamma A + (I_n-\Gamma)
\end{bmatrix},\qquad
N(\Gamma)=
\begin{bmatrix}
0\\ \Gamma
\end{bmatrix}.
\]
We begin the proof by defining a Lyapunov candidate $V(k)$ as
\begin{align*}
   V(k):=\max _{i, j}\left\|\eta_i(k)-\eta_j(k)\right\|_{\infty}
\end{align*}
Then, It follows immediately that $x_{i}(k) = x_{j}(k)$ for all $i,j \in \mathcal{V}$, whenever $V(k) =0$. Therefore, our objective is to prove that \(V(k) \to 0\) almost surely.

(ii) From \citep[Theorem 1.1]{hartfiel2006markov} applied coordinatewise and the
fact that $M(\Gamma)$ is row-stochastic, we obtain
\begin{align*}
    V(k+1)
    &\le \tau\big(M(\Gamma(k+1))\big)\,V(k) \\
     & + \max_{i,j}\big\|v_i(k+1)-v_j(k+1)\big\|_\infty,
\end{align*}
where $\tau(M):=1-\min_{r,s}\sum_j \min\{M_{rj},M_{sj}\}\in[0,1]$ is the
coefficient of ergodicity. By iteration and using that
$\prod_{s=1}^{t} M(\Gamma(k+s))$ remains row-stochastic, we obtain
\begin{align}
    V(k+t)
    &\le \tau\!\left(\prod_{s=1}^{t}M(\Gamma(k+s))\right)V(k) \notag\\
    &\quad + \sum_{s=1}^{t}\max_{i,j}\big\|v_i(k+s)-v_j(k+s)\big\|_\infty.\label{eq:proof1}
\end{align}

Since $\mathcal{G}$ is a connected graph and $1-d_i >0$, the matrix $W:= I_n-D + A$ is primitive \citep[Corollary 8.5.8]{horn2012matrix}. Therefore,  there exist an $L$ such that $W^{L-1}$ has all entries positive.
The $(i,j)$ entry of $W^{L-1}(I_n-D)$ equals 
\begin{align*}
    (W^{L-1}(I_n-D))_{ij}
\;=\; (W^{L-1})_{ij}\, (1-d_j).
\end{align*}
Since $(W^{L-1})_{ij}>0$ and $1-d_j>0$ for all $i, j \in \mathcal{V}$, there exists
$\epsilon>0$ such that $\big(W^{L-1}(I_n-D)\big)_{ij}\ge \epsilon$. Since we have 
\begin{align*}
    M(I_n)^{L} = \begin{bmatrix}
       W^{L-1}(I_n-D) & W^{L-1} A\\ W^{L-1}(I_n-D) & W^{L-1}A
    \end{bmatrix},
\end{align*}it follows that $ \sum_{j = 1}^{N}\min \left\{M(I_n)_{r j}, M(I_n)_{s j}\right\} > 0$ for all $r, s \in \mathcal{V}$. It further implies that
\begin{align}
\label{eq:proof2}
    \tau(M(I_n)^{L})\le 1-\epsilon<1.
\end{align}

Now we consider the event
$$
\mathcal{E}_k:=\{\Gamma(t)=I_n \text { for } t=k+1, \ldots, k+L\} .
$$
The probability of this event  can be calculated as,
$$
p_{\mathrm{event}} = \mathbb{P}\left(\mathcal{E}_k\right)=\prod_{t=k+1}^{k+L} \prod_{i=1}^N p_i > 0.
$$
Leveraging~\eqref{eq:proof1} and~\eqref{eq:proof2}, the conditional expectation of $V(k + L)$ given $\eta(k)$ can be bounded as follows:
\begin{align*}
    &\bE[V(k+L) \mid \eta(k)] \\ 
    &\leq \mathbb{P}\left(\mathcal{E}_k\right) \tau(M(I_n)^L)V(k) +  (1 - \mathbb{P}\left(\mathcal{E}_k\right) )V(k) \\
    &\qquad + \sum_{s=1}^{L}\mathbb{E}[\max_{i,j}\big\|v_i(k+s)-v_j(k+s)\big\|_\infty] \\
    &\leq (1-\epsilon p_{\mathrm{event}})V(k)  \\
    &\qquad + \sum_{s=1}^{L}\mathbb{E}[\max_{i,j}\big\|v_i(k+s)-v_j(k+s)\big\|_\infty]
\end{align*}
Using the bound on the expectation of the maximum of Gaussian random variables \citep[Proposition 2.7.6]{vershynin2018high}, it follows that
\begin{align*}
    \mathbb{E}\left[\max_{i,j}\big\|v_i(k+s)-v_j(k+s)\big\|_\infty \right] \leq 2 \sigma_{\max }(k)\sqrt{\log 2N},
\end{align*}
where $\sigma_{\max}(k) = \max \{\sigma_1(k), \dots, \sigma_{N}(k)\}. $ Therefore, we have 
\begin{align*}
    \bE[V(k+L) \mid \eta(k)] \leq (1-\epsilon p_{\mathrm{event}})V(k) +  \beta(k),
\end{align*}
where $\beta(k) : =\sum_{s=0}^{L-1}2 \sigma_{\max }(k+s)\sqrt{2 \log N}$. Since $\sigma_i(k) \to 0$ for all $i \in \mathcal{V}$, $\beta(k) \to 0$ also holds. 

(iii) Combining $\beta(k) \to 0$ with \(1 - \epsilon p_{\mathrm{event}} < 1\) and invoking \citep[Lemma A.1]{li2010consensus}, we conclude that
\begin{align*}
    V(k) \to 0 \quad \text{a.s.},
\end{align*}
which completes the proof.
\qed

\bibliography{ref}             
\end{document}